# Assessment of Developmental Dysgraphia Utilising a Display Tablet *

Jiri Mekyska[1[0000−0002−6195−193X]], Zoltan Galaz[1[0000−0002−8978−351X]], Katarina Safarova[2[0000−0003−0216−2057]], Vojtech Zvoncak[1[0000−0002−1948−4653]], Lukas Cunek[2], Tomas Urbanek[2[0000-0002-8807-4869]], Jana Marie Havigerova[2], Jirina Bednarova[2], Jan Mucha[1], Michal Gavenciak[1], Zdenek Smekal[1[0000−0002−8483−5448]], and Marcos Faundez-Zanuy[3[0000−0003−0605−1282]]

[1] Department of Telecommunications, Faculty of Electrical Engineering and Communication, Brno University of Technology, Brno, Czech Republic
mekyska@vut.cz
[2] Department of Research Methodology, Institute of Psychology, The Czech Academy of Sciences, Brno, Czech Republic
[3] Tecnocampus, Universitat Pompeu Fabra, Mataro, Barcelona, Spain

**Abstract.** Even though the computerised assessment of developmental dysgraphia (DD) based on online handwriting processing has increasing popularity, most of the solutions are based on a setup, where a child writes on a paper fixed to a digitizing tablet that is connected to a computer. Although this approach enables the standard way of writing using an inking pen, it is difficult to be administered by children themselves. The main goal of this study is thus to explore, whether the quantitative analysis of online handwriting recorded via a display/screen tablet could sufficiently support the assessment of DD as well. For the purpose of this study, we enrolled 144 children (attending the 3rd and 4th class of a primary school), whose handwriting proficiency was assessed by a special education counsellor, and who assessed themselves by the Handwriting Proficiency Screening Questionnaires for Children (HPSQ–C). Using machine learning models based on a gradient-boosting algorithm, we were able to support the DD diagnosis with up to 83.6 % accuracy. The HPSQ–C total score was estimated with a minimum error equal to 10.34 %. Children with DD spent significantly higher time in-air, they had a higher number of pen elevations, a bigger height of on-surface strokes, a lower in-air tempo, and a higher variation in the angular velocity. Although this study shows a promising impact of DD assessment via display tablets, it also accents the fact that modelling of subjective scores is challenging and a complex and data-driven quantification of DD manifestations is needed.

* This study was supported by a project of the Technology Agency of the Czech Republic no. TL03000287 (Software for advanced diagnosis of graphomotor disabilities) and by Spanish grant of the Ministerio de Ciencia e Innovaci´on no. PID2020-113242RB-I00.

## 1   Introduction

Handwriting is a complex perceptual-motor skill combining precise graphomotor movements, visual perception, visual-motor coordination, motor planning and execution, kinesthetic feedback, and orthographic coding [29]. It is a crucial skill that children acquire during their early years of schooling. Typically, around the ages of 8 to 10 [39], after 3 – 4 years of education and letter formation practice, handwriting becomes automatic, and children effortlessly and accurately produce letters.

Developmental dysgraphia (DD) refers to a condition where children experience difficulties in acquiring proficient handwriting skills, despite having normal cognitive abilities, ample learning opportunities, and an absence of neurological issues [7, 10, 25]. The occurrence of DD varies between countries, assessment methods, and raters, with prevalence rates ranging from 7 % to 34 % [21, 27]. Furthermore, studies have shown that boys tend to be diagnosed with DD more frequently than girls [19, 36]. DD can have a detrimental impact on various aspects of a child's daily life. This includes lower self-esteem, poor emotional well-being, as well as problematic communication and social interaction. In order to provide timely and effective therapy, and enhance the quality of life of children with DD, psychologists, special education counsellors, and other experts need a robust framework that enables accurate diagnosis and assessment.

Nowadays, psychologists or special education counsellors assess DD mainly subjectively using scales such as Handwriting Proficiency Screening Questionnaire (HPSQ) [30], Handwriting Legibility Scale (HLS) [5], or the shortened version of the Concise Assessment Methods of Children Handwriting (SOS: BHK) [38]. Moreover, some scales, such as Handwriting Proficiency Screening Questionnaires for Children (HPSQ–C) [33], were developed for children's self-evaluation. Nevertheless, approaches based on these scales could have several limitations, e.g. they are subjective, they rely on the perceptual abilities of a rater, they do not provide a complex assessment of the product/process of handwriting, etc.

One possible way of overcoming these limitations is to use decision-support systems that process online handwriting recorded by digitizing tablets [11, 26]. Such technology proved to bring interesting results and insights in both binary diagnosis [3, 14] and assessment [26, 40] of DD or graphomotor difficulties. For a comprehensive review, we refer to [20]. In all these studies the children performed handwriting/drawing tasks on a paper using a special inking pen. In a few works, the authors utilised protocols, where children wrote on displays/screens using plastic nibs (this way of writing has already been proven to be different when compared to writing on a paper [1]). In 2022, Asselborn et al. employed the Apple iPad to overcome the conventional binary diagnosis procedure and to assess handwriting difficulties on a scale, from the lightest cases to the most severe [2]. This revolutionary approach provided a global score, as well as four specific scores for kinematics, pressure, pen tilt and static features. The authors also highlighted that although two children could be diagnosed with DD, they could have different manifestations (e.g. kinematic vs. spatial). In the same year, Dui et al. introduced an Android app for handwriting skill screening at the preliteracy stage [15]. Using the Samsung Galaxy Tab A, the authors proved that the app, and generally drawing on a screen, could facilitate the detection of graphomotor disabilities in children attending a kindergarten. Lomurno et al. used the same app in a longitudinal study, where the authors monitord the development of graphomotor/handwriting skills in children starting in the last year of kindergarten and ending in the second class of a primary school [22]. In a sample of 210 children, they predicted a risk of dysgraphia with 84.62 % accuracy.

To the best of our knowledge, the three above-mentioned studies are the only ones where the authors used screen/display tablets to quantitatively analyse handwriting or graphomotor difficulties. Thus although writing/drawing on display tablets has increasing popularity, this relatively new field still contains some knowledge gaps. Asselborn et al. demonstrated that a display tablet could be used to assess the severity of DD [2]. Nevertheless, for that approach, they developed their own scale (in a data-driven way). The main goal of this study is to explore, whether the quantitative analysis of online handwriting recorded via a display tablet could sufficiently emulate children's self-assessment (which has not been explored before), and how well it models the diagnosis made by special education counsellors.

## 2 Materials and methods

### 2.1 Dataset

For the purpose of this study, we enrolled 62 children (12 intact girls, 23 intact boys, 13 girls with DD, 14 boys with DD; age = $9.2 \pm 0.5$ years) and 82 children (12 intact girls, 15 intact boys, 12 girls with DD, 43 boys with DD; age = $10.2 \pm 0.5$ years) attending the 3rd and 4th class of a Czech primary school, respectively. The children were assessed by a special education counsellor who stratified them into two groups: intact and dysgraphic. In addition, children assessed themselves by the Czech version of the HPSQ–C questionnaire [35], more specifically, they addressed 10 items (on a 5-point Likert scale ranging from 0 to 4, where a higher value means worse performance) that could be grouped into three factors: legibility (max value = 12), performance time (max value = 12), and physical and emotional well-being (max value = 16) [33, 35]. An overview of the HPSQ–C scores in both classes could be found in Table 1.

During the acquisition, the children were asked to perform a paragraph copy task on a display tablet Wacom Cintiq 16 (DTK1660K0B) using a stylus with a felt nib. Before the acquisition of this task, the children had time to get familiar with the tablet, e.g., by drawing a random picture. The content of the paragraph was selected depending on the class a child attended. Regarding the 3rd class,

**Table 1.** Overview of the HPSQ–C scores in both cohorts.

| | 3rd class | | | | | | |
|---|---|---|---|---|---|---|---|
| Score | Mean | Std | Min | Q1 | Median | Q3 | Max |
| legibility | 2.98 | 2.37 | 0 | 1 | 2 | 5 | 11 |
| performance time | 5.47 | 2.18 | 2 | 4 | 5 | 7 | 11 |
| well-being | 3.87 | 2.75 | 0 | 2 | 4 | 6 | 10 |
| total | 12.31 | 5.66 | 4 | 7 | 11.5 | 17 | 29 |
| | 4th class | | | | | | |
| Score | Mean | Std | Min | Q1 | Median | Q3 | Max |
| legibility | 3.71 | 2.20 | 0 | 2 | 3 | 5 | 10 |
| performance time | 5.66 | 2.26 | 1 | 4 | 6 | 7 | 11 |
| well-being | 5.05 | 2.91 | 0 | 2 | 5 | 7 | 12 |
| total | 14.41 | 4.96 | 5 | 11 | 14 | 18 | 27 |

children were asked to copy the following text (printed using capital letters) on the display using cursive letters:

*Maminčina třešňová marmeláda je nejsladší.*
*Nejoblíbenějším pamlskem našich návštěvníků jsou škvarky.*
*Babiččiny rozbité hodiny netikají.*

Examples of this task performed by an intact boy and a boy diagnosed with DD could be seen in Figure 1. Children attending the 4th class followed the same instructions, but copied the following text:

*Uprostřed náměstí se tyčil stříbrný sloup.*
*Pod střechou babiččiny chaloupky se uhnízdila vlaštovčí rodinka.*
*Tetička mi dala pestrobarevné odstřižky látek.*
*Nejbližší tramvajová zastávka je u víceúrovňové křižovatky.*

Parents of all children enrolled into this study signed an informed consent approved by the Ethics Committee of the Institute of Psychology of the Czech Academy of Sciences. Throughout the whole study, the Ethical Principles of Psychologists and Code of Conduct released by the American Psychological Association [4] were followed.

### 2.2 Feature extraction

The handwriting was sampled with frequency $f_s$ = 200 Hz and represented by a set of time-series: x and y position; timestamp; a binary variable, being 0 for the in-air movement (recorded up to 1.5 cm above the tablet's surface) and 1 for the on-surface one; pressure exert on the tablet's surface; pen tilt; and azimuth.

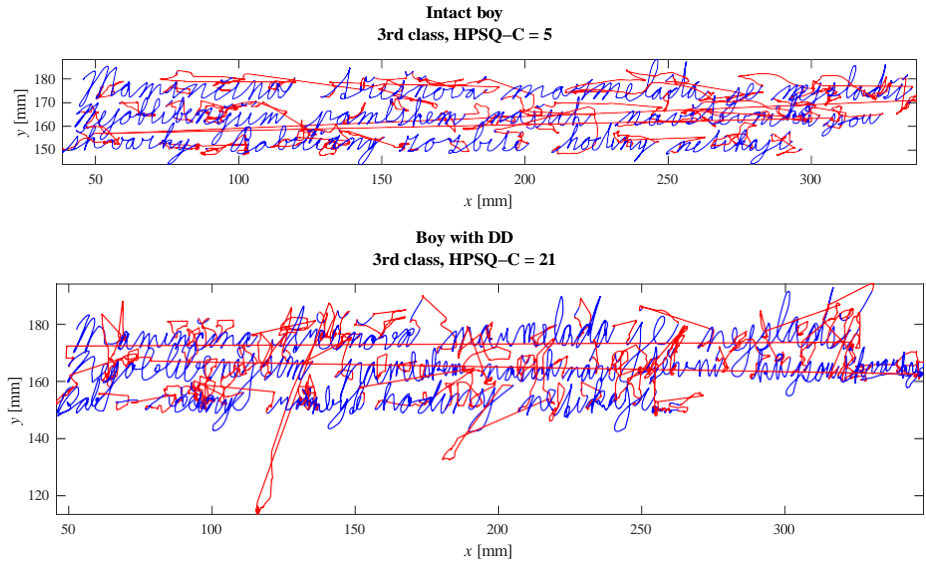

**Fig. 1.** Examples of the paragraph copy task performed by an intact boy and a boy diagnosed with DD (blue colour represents the on-surface movement, red colour the in-air one).

Consequently, these online handwriting signals were processed by the freely available Python library handwriting-features (v 1.0.5) [17]. More specifically, we extracted these conventional features [3, 14, 20, 27]:

1. temporal – duration of writing, ratio of the on-surface/in-air duration, duration of strokes, and ratio of the on-surface/in-air stroke duration
2. kinematic – velocity, angular velocity [24], and acceleration
3. dynamic – pressure, tilt, and azimuth
4. spatial – width and height of strokes
5. other – number of interruptions (pen elevations), relative number of interruptions, number of pen stops [28], tempo (number of strokes normalised by duration), Shannon entropy [8]

In this study, we consider the stroke as an on-surface/in-air trajectory between two pen elevations. Since some of these features are represented by a vector, we transformed them into a scalar value using statistics such as median, ncv – non-parametric coefficient of variation (defined as the median divided by the inter-quartile range), 95p – 95th percentile, and slope. The Shannon entropy, tempo, kinematic, and temporal features were calculated from the in-air movement as well. In addition, the Shannon entropy, velocity and acceleration were considered globally, but also in the horizontal/vertical projection.

### 2.3  Statistical analysis and machine learning

Since the dataset is not balanced in terms of sex, before any further processing, we have regressed out [37] this confounding factor from the feature values. Next, in order to get a first insight into the features' discrimination power, we calculated the Mann-Whitney U test comparing the intact and dysgraphic groups. To have an intuition whether a feature has generally higher/lower value in the intact group, we calculated Spearman's correlation between the feature values and the diagnosis performed by the special education counsellor. The same correlation was employed to explore the relations between feature values and the HPSQ–C scores (i.e. the total score and those evaluating the legibility, performance time and physical and emotional well-being). In this exploratory statistical analysis, the $p$ values were adjusted using the FDR (false discovery rate) correction [6]. The significance level was set to $a = 0.05$.

Next, we built binary classification (modelling the diagnosis) and regression (modelling the HPSQ–C scores) models using XGBoost algorithm [9]. The classification test performance was evaluated using balanced accuracy (BACC), Matthew's correlation coefficient (MCC), sensitivity (SEN), and specificity (SPE). The regression test performance was evaluated using mean absolute error (MAE), mean squared error (MSE), root mean squared error (RMSE), and estimation error rate (EER). EER is defined as MAE normalised by the theoretical range of values in the given score (to provide the error in terms of percentage).

In both cases, we optimized the models' hyperparameters using 500 iterations of randomized search strategy via stratified 10-fold cross-validation with 10 repetitions. The following hyperparameters were optimised: the learning rate [$0.001, 0.01, 0.1, 0.2, 0.3$], $\gamma$ [$0, 0.05, 0.10, 0.15, 0.20, 0.25, 0.5$], the maximum tree depth [$6, 8, 10, 12, 15$], the fraction of observations to be randomly sampled for each tree (subsample ratio) [$0.5, 0.6, 0.7, 0.8, 0.9, 1.0$], the subsample ratio for the features at each level [$0.4, 0.5, 0.6, 0.7, 0.8, 0.9, 1.0$], the subsample ratio for the features when constructing each tree [$0.4, 0.5, 0.6, 0.7, 0.8, 0.9, 1.0$], the minimum sum of the weights of all observations required in a child node [$0.5, 1.0, 3.0, 5.0, 7.0, 10.0$], and the balance between positive and negative weights [$1, 2, 3, 4$]. Finally, the models were interpreted via the SHAP (SHapley Additive exPlanations) [23] values of the top ten features.

## 3  Results

The results of the Mann-Whitney U test and Sperman's correlation with the diagnosis are reported in Table 2 (in each class we show the top 5 most discriminative features). In both cases, these top 5 features were significant ($p < 0.05$) even after the FDR correction. The most significant feature is the in-air movement duration, where children diagnosed with DD reach significantly higher values. The global duration plays a significant role as well (with the same direction). Moreover, the ratio of the on-surface/in-air movement duration suggests that children with DD spent significantly higher time in-air than on-surface. Children with DD also manifested a higher number of pen elevations, a bigger height of on-surface strokes, a lower in-air tempo, and a higher variation in the angular velocity.

**Table 2.** Results of the exploratory statistical analysis (the top 5 features).

| Feature | $p$ (MW) | $\hat{p}$ (MW) | $\rho$ (DG) | $p$ (DG) | $\hat{p}$ (DG) |
|---|---|---|---|---|---|
| *3rd class* | | | | | |
| duration of writing (in-air) | 0.0001 | 0.0043 | 0.51 | 0.0000 | 0.0017 |
| duration of writing | 0.0001 | 0.0043 | 0.49 | 0.0000 | 0.0017 |
| median height of stroke (on-surface) | 0.0002 | 0.0044 | 0.48 | 0.0001 | 0.0019 |
| number of interruptions | 0.0007 | 0.0104 | 0.44 | 0.0004 | 0.0062 |
| ratio of on-surface/in-air duration | 0.0008 | 0.0104 | -0.43 | 0.0005 | 0.0062 |
| *4th class* | | | | | |
| duration of writing (in-air) | 0.0001 | 0.0056 | 0.42 | 0.0001 | 0.0030 |
| ratio of on-surface/in-air duration | 0.0002 | 0.0056 | -0.42 | 0.0001 | 0.0030 |
| duration of writing | 0.0007 | 0.0162 | 0.38 | 0.0005 | 0.0111 |
| tempo (in-air) | 0.0013 | 0.0213 | -0.36 | 0.0009 | 0.0157 |
| ncv of angular velocity (on-surface) | 0.0022 | 0.0294 | 0.34 | 0.0017 | 0.0230 |

[1] $p$ – $p$ value; $\hat{p}$ – $p$ value after the FDR correction; $\rho$ – Spearman's correlation coefficient; MW – results of the Mann-Whitney U test; DG – Spearman's correlation with the diagnostic value

The results of the correlation analysis could be found in Table 3 and Table 4. In this case, for each HPSQ–C sub-score and the total score, we report the top 3 most significant features. Except for the correlation with the total score in the 4th class, none of the results passed the FDR correction, thus they must be considered critically. The legibility sub-score positively correlated with, e.g., increasing duration of in-air strokes (3rd class) or increased height of on-surface strokes (4th class). In terms of the performance time, we observed positive correlations with, e.g. higher maximum on-surface velocity/acceleration (3rd class) or increased duration (4th class). Regarding physical and emotional well-being, this sub-score positively correlated with, e.g., increased values of kinematic features (in both classes). Finally, concerning the HPSQ–C total score, we observed positive correlations with, e.g., higher on-surface strokes or higher values of vertical on-surface velocity (in both classes).

The results of the classification analysis are reported in Table 5. In the 3rd class, a model performed the diagnosis of DD with 74.3 % balanced accuracy (SEN = 88.6 %, SPE = 60.0 %), while in the 4th class, the model reached BACC = 83.6 % (SEN = 92.7 %, SPE = 74.6 %). The associated SHAP values could be found in Figure 2. The top 3 most important features in the first case were the

8      J. Mekyska et al.Table 3. Results of the correlation analysis (3rd class).

| Legibility | ρ | p | p̂ |
|---|---|---|---|
| slope of duration of strokes (in-air) | 0.29 | 0.0238 | 0.7767 |
| ncv of altitude | 0.29 | 0.0247 | 0.7767 |
| ncv of duration of pen stops | -0.26 | 0.0377 | 0.7767 |
| Performance time | ρ | p | p̂ |
| median height of strokes (on-surface) | 0.40 | 0.0012 | 0.0832 |
| 95p of vertical velocity (on-surface) | 0.33 | 0.0085 | 0.2749 |
| 95p of vertical acceleration (on-surface) | 0.31 | 0.0152 | 0.2749 |
| Well-being | ρ | p | p̂ |
| 95p of vertical acceleration (on-surface) | 0.33 | 0.0094 | 0.3550 |
| 95p of vertical velocity (on-surface) | 0.31 | 0.0137 | 0.3550 |
| median of vertical velocity (on-surface) | 0.31 | 0.0157 | 0.3550 |
| Total | ρ | p | p̂ |
| ncv of duration of pen stops | -0.32 | 0.0126 | 0.5101 |
| 95p of vertical velocity (on-surface) | 0.28 | 0.0253 | 0.5101 |
| median height of strokes (on-surface) | 0.28 | 0.0274 | 0.5101 |

[1] $\rho$ – Spearman's correlation coefficient; $p$ – $p$ value; $p\hat{}$ – $p$ value after the FDR correction

slope of the duration of on-surface strokes, the slope of horizontal in-air velocity, and the slope of pressure (the directions suggest that DD children had increased slope in all three features). Regarding the 4th class, the top 3 most important features were the slope of angular on-surface velocity (higher in the DD group), the ratio of the on-surface/in-air movement duration (lower in the intact group), and the variation of angular on-surface velocity (lower in the intact group).

Results of the regression analysis could be found in Table 6. The legibility sub-score was estimated with 17.96 % and 14.90 % error in the 3rd and 4th class, respectively. In terms of the performance time, we reached 16.48 % (3rd class) and 16.31 % (4th class) errors. Regarding the physical and emotional well-being, this sub-score was estimated with 15.34 % and 16.69 % error. The HPSQ–C total score was in the 3rd class estimated with 14.00 % error (MAE = 5.60) and in the 4th class with an error equal to 10.34 % (MAE = 4.13). The SHAP values of models estimating the total score are shown in Figure 3. E.g., in the 3rd class, the model mostly relied on the 95p of vertical in-air velocity (higher in the DD group), the variation of horizontal in-air velocity (lower in the intact group), and the 95p of vertical on-surface acceleration (higher in the DD group). In the 4th class, the most important features were the 95p of vertical/horizontal/global on-surface acceleration (higher in the DD group).

Table 4. Results of the correlation analysis (4th class).

| Legibility | $\rho$ | $p$ | $\hat{p}$ |
|---|---|---|---|
| median height of strokes (on-surface) | 0.36 | 0.0008 | 0.0557 |
| ratio of on-surface/in-air duration | -0.32 | 0.0035 | 0.0936 |
| 95p of vertical velocity (on-surface) | 0.31 | 0.0041 | 0.0936 |
| Performance time | $\rho$ | $p$ | $\hat{p}$ |
| median of angular velocity (on-surface) | -0.23 | 0.0382 | 0.6137 |
| duration of writing | 0.22 | 0.0461 | 0.6137 |
| ncv of duration of strokes (in-air) | 0.22 | 0.0489 | 0.6137 |
| Well-being | $\rho$ | $p$ | $\hat{p}$ |
| median of vertical velocity (on-surface) | 0.33 | 0.0025 | 0.0871 |
| median of velocity (on-surface) | 0.33 | 0.0026 | 0.0871 |
| ncv of tilt | -0.30 | 0.0055 | 0.0963 |
| Total | $\rho$ | $p$ | $\hat{p}$ |
| median height of strokes (on-surface) | 0.35 | 0.0013 | 0.0363 |
| 95p of vertical velocity (on-surface) | 0.34 | 0.0020 | 0.0363 |
| 95p of velocity (on-surface) | 0.33 | 0.0022 | 0.0363 |

[1] $\rho$ – Spearman's correlation coefficient; $p$ – $p$ value; $\hat{p}$ – $p$ value after the FDR correction

Table 5. Results of the classification analysis.

| Class | BACC [%] | MCC | SEN [%] | SPE [%] |
|---|---|---|---|---|
| 3rd | 74.3 | 0.5068 | 88.6 | 60.0 |
| 4th | 83.6 | 0.6841 | 92.7 | 74.6 |

[1] BACC – balanced accuracy; MCC – Matthew's correlation coefficient; SEN – sensitivity; SPE – specificity

## 4  Discussion

The results of the exploratory analysis showed that children with DD spent significantly higher time in-air. This result is in line with findings of previous studies [3, 34] and could be probably explained by difficulties recalling a correct letter shape, which is linked with a poor orthographic coding process [27]. Children with DD also manifested higher strokes. Rosenblum et al. assume that this is the effect of trying to write legibly because oversized letters do not require so big precision of handwriting [32]. In addition, in accordance with other studies, we observed that DD children manifested a higher number of pen elevations [27, 31]. To sum up, even though in our study we acquired handwriting using a display tablet, the most discriminative features were identical or similar to those reported in studies employing the paper and digitizer setup. Nevertheless, in order to precisely explore differences in features that discriminate between intact children and children with DD who write on paper or display, we would need to have tasks performed by a child on both types of devices. This point deserves further attention in future studies and research.

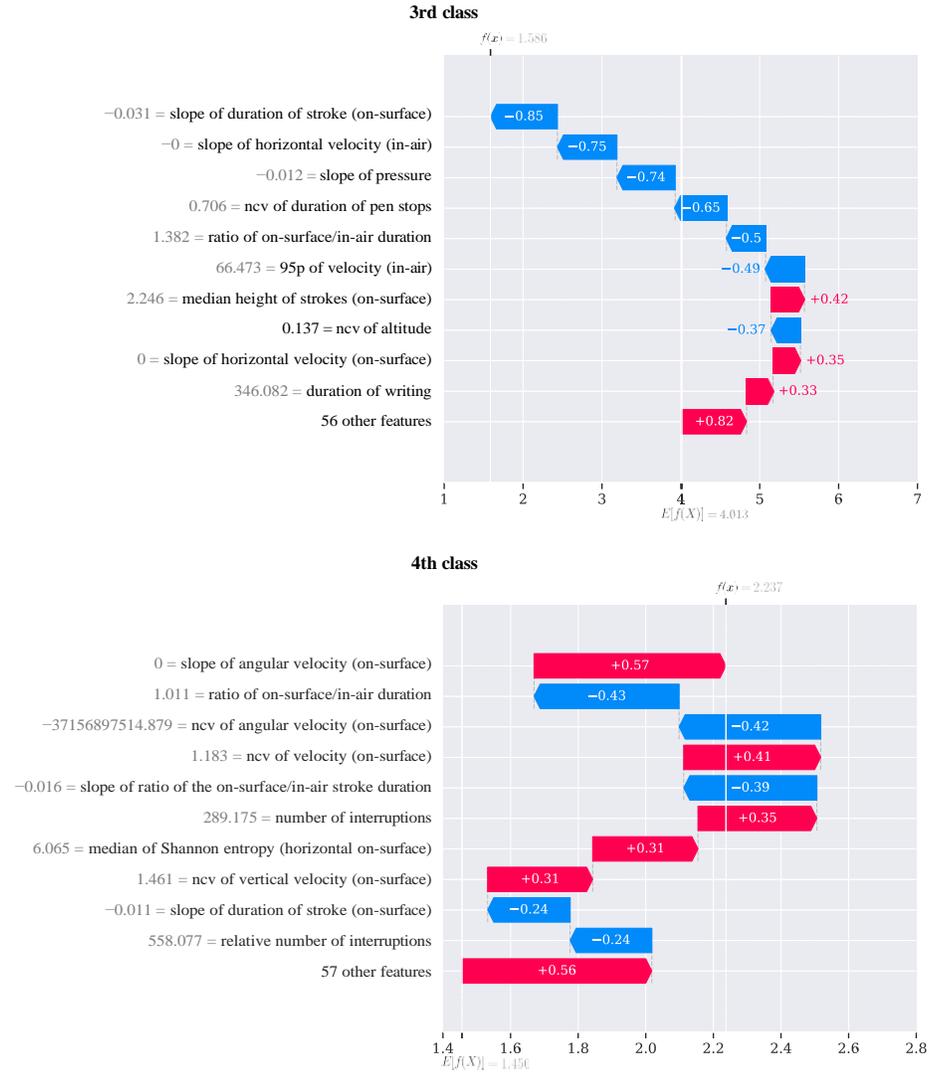

**Fig. 2.** SHAP values of the classification models.

An interesting finding was done when we correlated the features with the HPSQ–C sub-scores and with the total score. In many cases, we identified that the scores positively correlated with the median or 95th percentile of on-surface velocity meaning that children with the less proficient handwriting achieved higher velocities. On the one hand, this is against the findings of many studies (please see the review in [27]), on the other hand, a couple of teams reported that experts should rather focus on the speed-accuracy trade-off meaning that the low velocity does not have to be linked with the less proficient handwriting, but it could be associated with better accuracy of writing [16, 21].

Table 6. Results of the regression analysis (mean ± std).

| Score | MAE | MSE | RMSE | EER [%] |
|---|---|---|---|---|
| *3rd class* | | | | |
| legibility | 2.16 ± 0.40 | 7.23 ± 2.17 | 2.65 ± 0.43 | 17.96 ± 3.34 |
| performance time | 1.98 ± 0.44 | 6.48 ± 2.19 | 2.51 ± 0.45 | 16.48 ± 3.70 |
| well-being | 2.45 ± 0.46 | 8.90 ± 3.30 | 2.94 ± 0.53 | 15.34 ± 2.86 |
| total | 5.60 ± 1.10 | 44.23 ± 19.85 | 6.50 ± 1.41 | 14.00 ± 2.74 |
| *4th class* | | | | |
| legibility | 1.79 ± 0.31 | 5.31 ± 1.71 | 2.27 ± 0.37 | 14.90 ± 2.58 |
| performance time | 1.96 ± 0.42 | 6.09 ± 2.46 | 2.43 ± 0.46 | 16.31 ± 3.53 |
| well-being | 2.67 ± 0.36 | 10.46 ± 3.48 | 3.19 ± 0.51 | 16.69 ± 2.25 |
| total | 4.13 ± 0.18 | 27.35 ± 2.34 | 5.22 ± 0.23 | 10.34 ± 0.46 |

[1] MAE - mean absolute error; MSE – mean squared error; RMSE – root mean square error; EER – equal error rate

When performing the binary diagnosis, we achieved BACC = 74.3 % and BACC = 83.6 % in the 3rd and 4th class, respectively. These accuracies are not high and we think there is still a place for improvement. On the other hand, most of the available studies reported diagnostic accuracy in a range between 70 % and 90 % [13, 14, 18, 20] suggesting that we cannot fully rely on the scores provided by special education counsellors, because their assessment is subjective, inconsistent (e.g. one rater focuses more on the product and another one more on the process [27]), and with a questionable inter- and intra-rater variability [12]. Thus we believe that rather than performing a binary diagnosis and rather than modelling unreliable scores depending on human perception, we must introduce a concept that would be semi-data driven (i.e. less dependent on a human) and that would provide a detailed and objective assessment of manifestations associated with DD (because with this, we can even discriminate between several sub-types of dysgraphia and introduce a more focused therapy). For this purpose, we have developed the Graphomotor and Handwriting Disabilities Rating Scale (GHDRS), a first scale of its kind addressing the above-mentioned requirements [27].

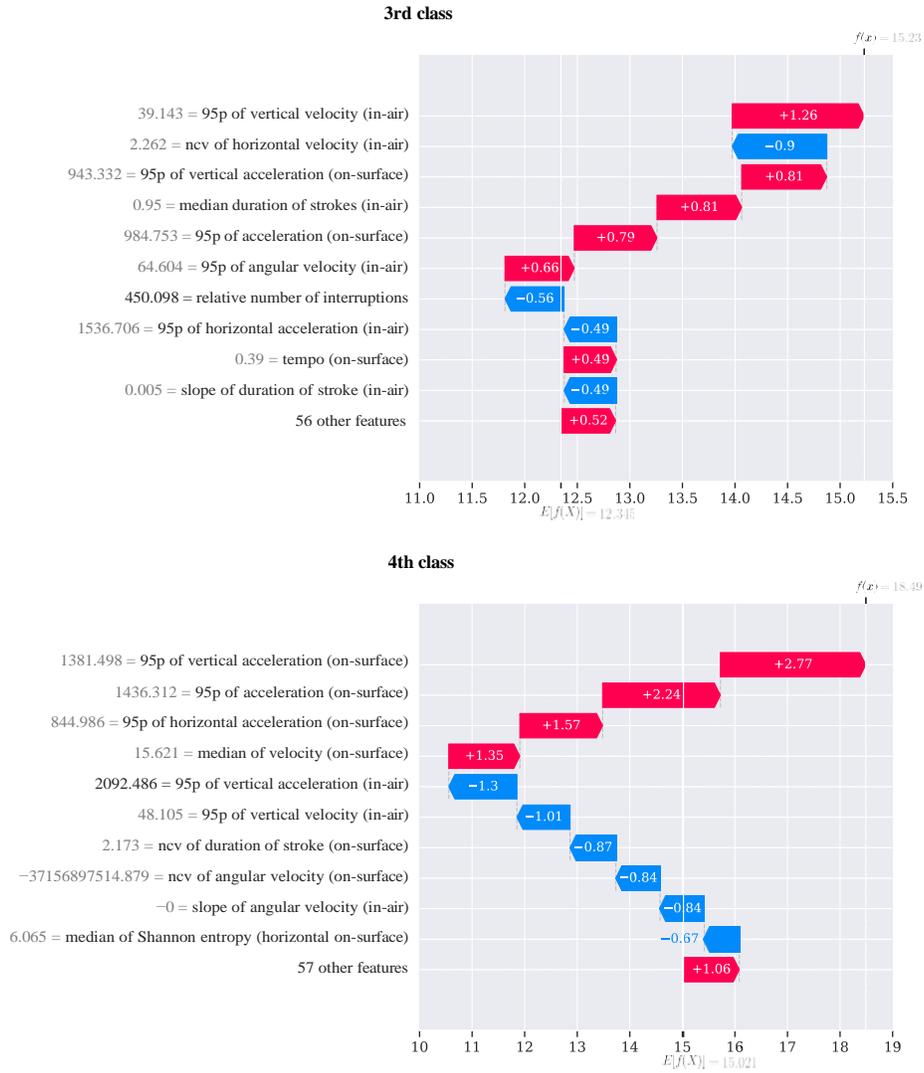

**Fig. 3.** SHAP values of the regression models.

Finally, we explored whether display tablets could be used to emulate children's self-assessment. The errors in the sub-scales ranged between 14.90 % and 17.96 %. The HPSQ–C total score was estimated with 14.00 % error in the 3rd class and with 10.34 % error in the 4th class. This is close to the results reported by Zvoncak et al. [40], however, who used the paper and digitizer setup.

## 5   Conclusion

The aim of this study was to explore, whether the easy-to-administer acquisition of online handwriting via display tablets could be used for a supportive diagnosis and assessment of DD. Our findings suggest that this approach could provide results comparable to those based on the paper and digitizer setups. Even though we used a display tablet connected to a laptop, which is still less comfortable, we proved that writing on a display/screen has a good potential in DD screening and further steps, such as transfer to e.g. iPad or Samsung Galaxy Tab technologies, could make the whole process even more comfortable and easy-to-use. In addition, we recommend using scales less dependent on human perception, while providing a complex overview of manifestations associated with DD (e.g. the GHDRS scale).

This study has several limitations. Firstly, the experiments were conducted on a database with a relatively small sample size. In order to generalize the conclusions, further studies should be performed. Additionally, although the effect of sex was regressed out, a more in-depth analysis of the impact of this confounding factor would be beneficial. Lastly, the diagnostic scores were provided by only one special education counselor. Having multiple raters would enhance the reliability of these scores.

16      J. Mekyska et al.